\documentclass[runningheads]{llncs}
\usepackage[T1]{fontenc}
\usepackage{amsmath}
\usepackage{graphicx}
\usepackage{verbatim}
\usepackage{hyperref}
\begin{document}
%
\title{Verifiable Randomness for Blockchain-Based Lottery Systems}
%
%
\author{Gonçalo Ferreira\inst{1,3}\orcidID{0009-0006-1029-9490} \and
André Zúquete\inst{2,3}\orcidID{0000-0002-9745-4361} \and
Paulo Bartolomeu\inst{2,4}\orcidID{0000-0002-1975-6233}}
\authorrunning{G.Ferreira et al.}
%
\institute{
Univ. of Aveiro, Campus Univ. de Santiago, 3810-189 Aveiro, Portugal
\and
DETI, Univ. of Aveiro, Campus Univ. de Santiago, 3810-189 Aveiro, Portugal
\and
IEETA / LASI, Univ. of Aveiro, Campus Univ. de Santiago, 3810-189 Aveiro, Portugal
\and
Instituto de Telecomunicações, Campus Univ. de Santiago, 3810-189 Aveiro, Portugal
}
\maketitle              
\begin{abstract}
As lotteries and other high-stakes decentralized applications increasingly depend on unpredictable randomness for their operations, the lack of a secure and transparent on-chain random number generator that is verifiable by all participants remains a critical open problem. Various approaches to blockchain-based random number generation have emerged over the years, each with their own strengths and limitations, and have consistently been superseded as blockchain technology evolved. This paper surveys existing approaches to on-chain randomness and proposes a new platform that builds upon the well-known commit-and-reveal scheme while directly addressing its principal vulnerability, the last revealer attack, in which the final participant can withhold their reveal in order to bias or abort the output upon seeing an unfavorable result. We further compare this solution with prior approaches and evaluate its entropy properties. The proposed architecture combines a web-based front-end with a Solidity smart contract deployed on the Polygon~2.0 blockchain.  Implemented and tested on the Amoy testnet, the prototype is low-cost and simple to deploy, providing a practical, accessible proof-of-concept for verifiable on-chain randomness. 

\end{abstract}

\section{Introduction}
For centuries, humans have relied on unpredictable forces to make unbiased decisions~\cite{mlodinow08}. From weather-based rituals determining leadership to dice rolls settling medieval disputes~\cite{david62}, randomness has served as a substitute for human judgment when fairness is required.

This reliance on chance goes deeper than practical utility. When outcomes are uncertain or personal bias might distort judgment, people instinctively defer to external sources of randomness, not out of irresponsibility, but out of a desire for decisions that cannot be blamed on fallible reasoning.

\subsection{Motivation}
The problem arises when randomness shifts from convenience to critical infrastructure. When a random outcome determines who wins a ten-million-euro lottery prize, which validator controls the next blockchain block, or how limited NFTs are distributed among thousands of buyers, this is insufficient. Malicious actors have clear incentives to exploit the generation process: a casino operator who can bias a slot machine, a lottery administrator who can predict winning numbers, a blockchain validator who can skew random selection, each representing a catastrophic failure of fairness~\cite{bonneau15}.

The challenge is to design systems that provide not just fair randomness, but fairness that can be proven fair. Participants must be able to verify mathematically that no party, not system operators, not other users, not the verifiers themselves, could have manipulated the outcome. This requirement for cryptographically verifiable fairness drives the development of blockchain-based randomness solutions over trust-based alternatives.

\subsection{Contribution}
This proposes a platform for conducting verifiable random processes in which each participant contributes a share of the randomness. After the process completes, a cryptographic proof allows any observer to confirm that the result was not manipulated. Participant's contributions are managed by a Smart Contract and stored on a blockchain, making the platform and all its inputs publicly auditable. Corrupting the process requires corrupting the underlying blockchain network.

The paper is structured as follows.
Section~\ref{rw} presents related work on the use of blockchains to produce random outcomes.
Section~\ref{arch} presents the architecture of our lottery system, in particular the algorithm we designed to produce random outcomes.
Section~\ref{quality} presents some preliminary statistical evaluation of the randomness of the lottery results.
Section~\ref{security} makes a security analysis.
Finally, Section~\ref{conc} concludes the paper and draws some future research lines.

\section{Related Work on Decentralized Blockchain RNGs}
\label{rw}

In distributed settings, multiple distrustful parties must collectively produce unbiased outcomes without delegating trust to any single entity. Rabin's randomness beacon~\cite{rabin83} introduced the concept of a publicly verifiable source, though centralized. Blum's coin-flipping protocol~\cite{blum82} showed that two distrustful parties could jointly generate an unbiased bit, a construction Sweeney and Shamos~\cite{CMU-ISRI-04-126} generalized into the \textit{RandomSelect()} protocol for $n$ participants, achieving $O(n)$ communication complexity under the assumption that at least one participant behaves honestly.

The first blockchain application of on-chain randomness was SatoshiDice\footnote{\url{https://www.satoshidice.com}}, launched by Erik Voorhees in April 2012. The system combined a transaction hash with a server-side secret inside SHA-512; the first four bytes of the result determined the outcome. Publishing the secret's hash after each round introduced provably fair gaming to blockchain, yet the design remained partially trust-dependent: the server-side secret was withheld until after the fact, and miner influence over transaction hashes posed a residual manipulation risk.

Commit-reveal schemes, the concept formalized in ~\cite{brassard1988}, addressed the mining influence problem by distributing entropy among participants. Each participant commits to a private value by publishing its hash on-chain, then reveals it once all commitments are collected; the final output combines all disclosed values. RANDAO extends this construction with Verifiable Delay Functions (VDFs~\cite{boneh2018vdf}) to prevent last-revealer bias~\cite{do2024randao} and uses token-based staking to disincentivize withholding.The scheme's limitation is liveness: a delayed or withheld reveal degrades output quality~\cite{lee2025}, and its security model assumes that slashing penalties outweigh manipulation gains, an assumption that does not always hold~\cite{yakira2020}.

Verifiable Random Functions (VRFs), introduced by Micali, Rabin and Vadhan~\cite{rabin99}, offer a stronger construction: only the holder of a secret key can compute the output, while anyone with the public key can verify it. Uniqueness, pseudorandomness, and verifiability make VRFs well-suited to decentralized settings. Algorand applies VRFs at the consensus layer~\cite{algorand}, selecting committee members anonymously each round through a stake-weighted threshold evaluation. For smart contract platforms, Chainlink VRF~\cite{chainlinkvrf} implements a Goldberg VRF specification~\cite{rfc:9381} through a decentralized oracle network, becoming the most widely used source of on-chain randomness across multiple chains. Both designs retain off-chain dependencies~\cite{hassan2023}, motivating fully on-chain alternatives~\cite{chompurmpakdee2023}.

\begin{figure}[t]
    \centering
    \includegraphics[width=1\linewidth]{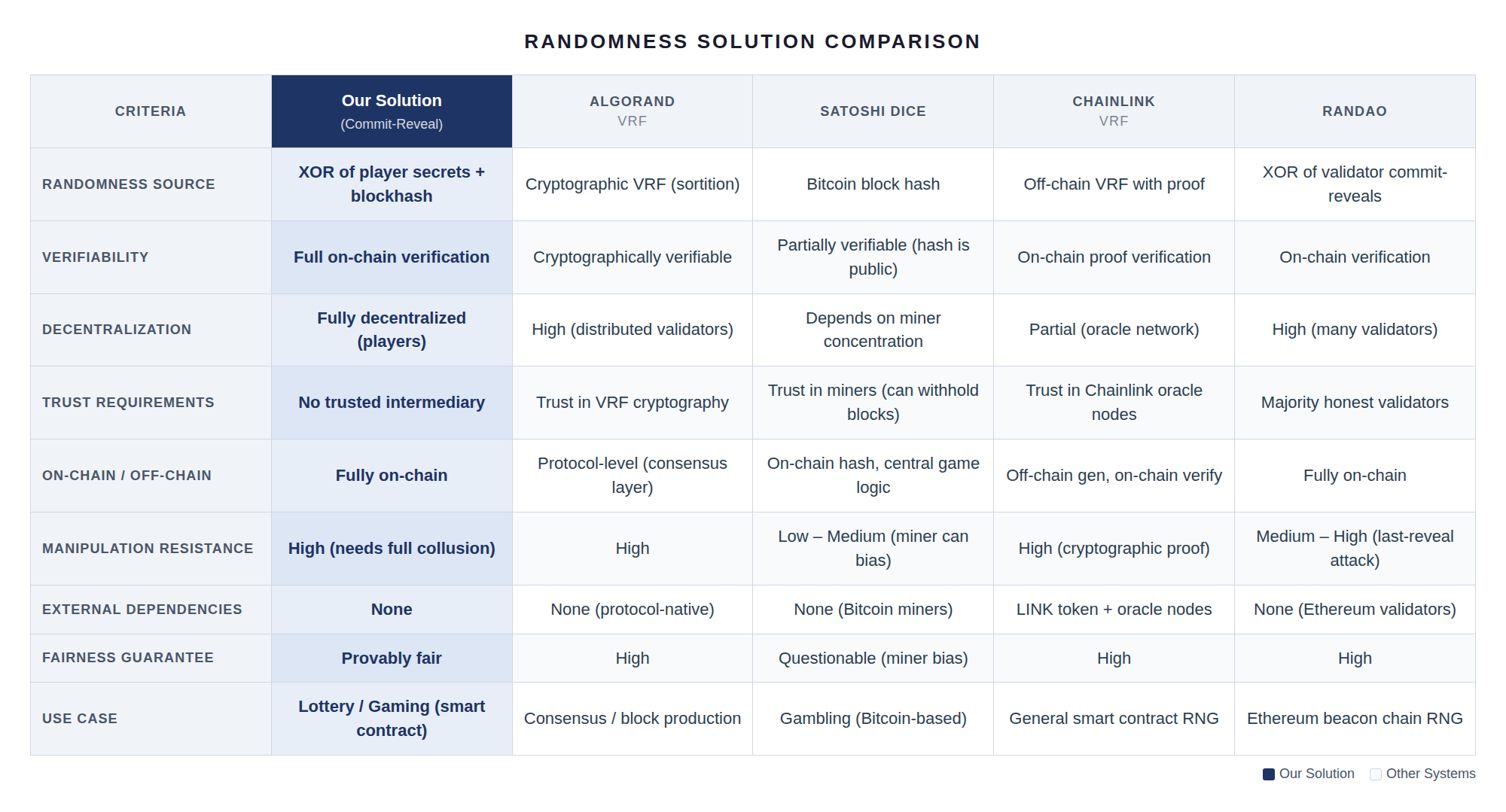}
    \caption{Comparison of randomness generation solutions across key evaluation criteria.}
    \label{fig:rng_comparison}
\end{figure}

As illustrated in Figure~\ref{fig:rng_comparison}, the five solutions differ substantially
across decentralization, trust requirements, and resistance to manipulation. Satoshi Dice
is the most rudimentary, deriving randomness from Bitcoin block hashes and remaining
vulnerable to miner bias. RANDAO improves on this through
validator commit-reveals but inherits the last-revealer attack as a structural
weakness. Algorand VRF eliminates this through cryptographic
.sortition, offering the strongest protocol-level guaranties, but is inseparable from
Algorand's consensus and cannot be used as a general-purpose primitive. Chainlink VRF
targets the general smart contract use case with off-chain VRF generation and on-chain
proof verification, though at the cost of oracle trust and recurring LINK token fees.

Our commit-reveal solution is fully on-chain, trustless, and free of external dependencies.
Randomness is derived from the XOR of player secrets combined with a future block hash,
making unilateral manipulation infeasible even with full participant collusion. Compared to
RANDAO, the last-revealer attack is treated in a different way, mixing methods to mitigate this attack;
compared to Chainlink VRF, the formal VRF guaranty is
traded for complete decentralization and zero recurring cost, a trade-off well-suited to
permissionless lottery applications on Ethereum Virtual Machine (EVM) compatible chains.

\section{Proposed Architecture}
\label{arch}

Following a review of existing approaches to random number generation
using blockchain technologies, we present an architecture for
cooperative, verifiable randomness in a lottery use-case. The proposed
system is built around a commit-reveal scheme with a mechanism we call
\textit{reveal-as-a-choice}, designed to address the last-revealer
problem without relying on external oracles or imposing heavy financial
penalties.

The platform covers all the core use-cases of a traditional lottery,
operating entirely on a public blockchain and structured around three
roles: the participant, the creator, and the smart contract itself,
treated as an autonomous on-chain actor. Participants can browse open
lotteries, enter either standard or password-protected rounds, view
the tickets held by their wallet, transfer or receive tickets between
wallets, and share lottery links with others. Creators configure new
rounds by setting parameters such as ticket price, duration, and prize
policy, choosing between public and password-protected access; the
latter additionally requires on-chain validation of a password hash
at entry. The smart contract executes the steps that must run without
human intervention, namely validating password hashes, drawing the
winner once a round closes, and distributing the prize to the winning
wallet, which underwrites the trustless guaranties of the design.

The main limitation inherent in any blockchain-based system is that
ticket purchases are not fully refundable; mechanisms for cancelable
lotteries or refundable tickets are left for future work. The current
focus is on providing a fully functional, fair, and trustless lottery
system that operates without reliance on third parties.

\subsection{System Overview}

The deployment on a public blockchain is a deliberate design choice that underpins the system's security guaranties. A public blockchain, such as Polygon~2.0, provides a transparent, immutable, and censorship-resistant execution environment, guaranteeing that every transaction and smart contract call is recorded in a globally replicated ledger that any party can independently audit.

Transparency is essential for a randomness-based application where any participant can verify that the lottery logic was executed exactly as specified, that no tickets were added or removed after commitment, and that the winning index was derived from the agreed-upon inputs without manipulation. Furthermore, the decentralized consensus mechanism ensures that no single entity, including the lottery deployer, can unilaterally alter the contract state or censor reveal transactions. The smart contracts are self-enforcing: once deployed, their rules cannot be changed without network consensus, eliminating the need to trust any intermediary. The system runs entirely on-chain through a set of smart contracts with no dependency on external randomness oracles, such as in Chainlink VRF.

A \texttt{LotteryFactory} contract uses the minimal proxy (clone) pattern to efficiently deploy individual \texttt{LotteryCloneable} instances, each representing a distinct lottery. The lottery creator configures parameters including the ticket price, maximum number of participants, duration of the commit phase, and optionally a reward for participants who reveal their committed values.

  The lottery lifecycle proceeds in three phases: \textit{entry} (commit), \textit{reveal},
  and \textit{resolution}.

  \paragraph{Entry Phase:}
  When entering a lottery, each participant selects a secret string and submits a
  cryptographic commitment to the smart contract. Specifically, the participant computes:
  \[
    r = \texttt{keccak256}(\textit{secret})
  \]
  \[
    c = \texttt{keccak256}(\textit{address} \mathbin\| r)
  \]
  where $c$ is the commitment hash stored on-chain. The actual secret is never transmitted;
  only its double-hash is recorded in \texttt{commits[msg.sender]}. Each participant also
  pays the ticket price in ETH and receives an ERC-721 NFT receipt as proof of participation.
  The entry phase ends when either the maximum number of participants is reached or the
  configured time limit expires.

  \paragraph{Reveal Phase:}
  After the entry phase closes, the lottery creator initiates the reveal phase by calling
  \texttt{startReveal(deadlineBlock)}, setting a specific future block number as the reveal
  deadline. During the reveal phase, participants call \texttt{reveal(r)}, submitting their
  original value $r$. The contract verifies the submission by recomputing the
  commitment
  \[
    \texttt{keccak256}(\textit{address} \mathbin\| r) \stackrel{?}{=} \texttt{commits[msg.sender]}
  \]
  If the check passes, the participant's revealed value replaces their stored commitment, and
  their participation in the seed computation is confirmed.

  \paragraph{Resolution:}
  Once the reveal deadline block has been mined, the lottery creator calls
  \texttt{announceWinner()}, which computes the winning seed and distributes the prize pool.
  This call must occur within 255 blocks after the deadline, due to the EVM constraint that
  only allows \texttt{blockhash()} queries within the most recent 256 blocks.

  \subsection{Seed Generation}

  The final randomness seed is computed deterministically from three entropy sources:

  \begin{enumerate}
      \item \textbf{Aggregated reveals.} The XOR of all stored \texttt{commits} values
      (which, after the reveal phase, correspond to either verified revealed values or
      the original commitment hashes of non-revealers):$$
        A = \bigoplus_{i=1}^{n} \texttt{commits}[P_i]
      $$
      \item \textbf{Block entropy.} The hash of the reveal deadline block:$$
        B = \texttt{blockhash}(\textit{revealDeadlineBlock})
      $$
      \item \textbf{Reveal count.} The number of participants who successfully revealed,
      preventing the seed from being reproducible without this information.
  \end{enumerate}

  The seed is then computed as:$$
    \text{seed} = \texttt{keccak256}(A \mathbin\| B \mathbin\| \textit{revealCount})
  $$
  and the winner is selected as \texttt{participants[seed \% n]}, where $n$ is the total
  number of participants.

\subsection{Reveal-as-a-Choice}

In classical commit-reveal schemes, all participants must reveal for the protocol to
complete, creating a denial-of-service vulnerability: any participant who dislikes an
emerging outcome can abort the lottery by withholding their reveal. Common
countermeasures require a security deposit slashed for non-revealers, adding friction
and financial risk for honest participants.

The proposed scheme takes a different approach: \textit{revealing is optional}. The
pre-committed hashes of non-revealers are included in the XOR aggregation regardless of
whether they reveal. A participant $P_i$ who does not reveal contributes with their
commitment hash $c_i = \texttt{keccak256}(\mathit{address}_i \mathbin\| r_i)$ to the
aggregate instead of $r_i$. Winner announcement executes unconditionally after the reveal deadline, satisfying the following liveness property:

\begin{quote}
\textbf{Liveness.} For any subset $S \subseteq \{P_1, \ldots, P_n\}$ of participants
who withhold their reveals, the lottery resolves and a winner is selected.
\end{quote}

\noindent
The entropy contribution of a non-revealer is not vacuous. Since
$c_{i}$ is computed from $r_i$, which
was chosen privately and is unknown to all other participants, $c_i$ is
computationally indistinguishable from uniform in the random oracle model.
Non-revealers therefore contribute with entropy to the seed at the same level as revealers,
from the perspective of all other participants. The only information asymmetry is that
$P_i$ knows both $r_i$ and $c_i$; however, since neither value can be changed after the
commit phase (by the binding property of $\mathcal{H}$), this knowledge does not confer
strategic advantage.

A further structural defense arises from the uncertainty surrounding the position of the last revealer. No participant can know in advance whether they will be the final one to reveal: this status is determined only ex post by the order in which reveal transactions are included on-chain, and is not observable until the reveal window closes. A participant who happens to be the last revealer is reduced to a binary choice, reveal correctly or withhold, with both branches still subject to the unknown $\texttt{blockhash}(\mathit{deadlineBlock})$ analyzed above. The strategic value of the last-revealer position is therefore bounded by the participant's ability to predict that they will occupy it, which is itself a low-probability event in any non-trivial lottery.

This residual case can be further mitigated by introducing a threshold parameter $t$ with
$1 \le t < n - 1$ limiting the maximum number of reveals for the lottery, and thus adding entropy to
the selection of the actual revealers that contribute to the lottery outcome. Lower $t$ values increase the competition of the participants for contributing as a revealer, which increases the degree of uncertainty and reduces the control of any participant to act as a last-revealer. A currency reward for revealers can serve as a further lever to promote competition among them.
Determining the optimal value of $t$ for a given participant count and adversarial model requires empirical evaluation across a range of configurations, which is left as future work.

Combined with the unknown deadline blockhash, these mechanisms (reveal-as-a-choice and the threshold $t$ for the maximum number of revealers) remove control of non-revealers while keeping the randomness of the outcome and eliminate denial-of-service
risk without requiring financial penalties or trust in any external enforcement
mechanism.

  \subsection{Mitigating the Last-Revealer Problem via Block Entropy}

  To neutralize this advantage, the system introduces an implicit $(n+1)$-th contributor:
  the hash of the reveal deadline block itself. When the lottery creator calls
  \texttt{startReveal(deadlineBlock)}, the chosen block has not yet been mined. Its hash is
  therefore unknown to all participants, including the last revealer, at the time they must
  decide whether to submit their reveal.
  
  The seed thus depends on an entropy value that is not available until after the reveal deadline passes.
Concretely, a participant considering whether to reveal in block $b < \textit{deadlineBlock}$
  cannot compute: $$
    \text{seed} = \texttt{keccak256}(A \mathbin\| \texttt{blockhash}(\textit{deadlineBlock})
    \mathbin\| \textit{revealCount})
  $$
 because \texttt{blockhash}(\textit{deadlineBlock}) is not yet known. This means that neither
  branch of their decision tree can be evaluated, eliminating the strategic advantage of
  waiting to be the last.

  The blockhash can be understood as a guaranteed final contributor $P_{n+1}$ whose value
  is determined by blockchain consensus after all reveal decisions have been made. Since
  miners would need to simultaneously control the reveal phase and the block mining process
  to exploit this, the attack surface is substantially reduced compared to RANDAO-style
  schemes.

  The constraint that \texttt{announceWinner()} must be called within 255 blocks of the
  deadline further ensures that the blockhash cannot be substituted or bypassed and that
  the resolution window is bounded and predictable.

\section{Early Results on Randomness Quality}
\label{quality}

The statistical quality of the proposed scheme was assessed through three complementary experiments: 
\begin{itemize}
\item An off-chain simulation of 256 lotteries with 10
participants each (XOR aggregation only, no deadline blockhash); and
\item Two on-chain runs
against a local Hardhat node executing the full commit-reveal-resolve life cycle:
Run~A (1{,}000 lotteries) and Run~B (10{,}000 lotteries), both with 5 participants, a 15-block reveal window, and zero failures.
\end{itemize}

\noindent
Across all runs, our scheme produced 256-bit seeds with no collisions and an average Hamming
weight within 0.13\,\% of the theoretical optimum. Rendering the 10{,}000 Run~B
seeds as a binary bitmap
showed no vertical stripes, horizontal bands, or diagonal grid patterns, making the output visually
indistinguishable from that produced by an MT19937 Mersenne Twister PRNG~\cite{Matsumoto98}.%

Winner-selection uniformity was confirmed by a $\chi^2$ goodness-of-fit test
(uniform $H_0$ over 5 outcomes, $df=4$, $\chi^2_{0.05,4}=9.49$). Run~A yielded
$\chi^2=1.01$ and Run~B yielded $\chi^2=1.03$, both far below the critical value,
so $H_0$ is not rejected in either case.
The Shannon entropy of winner
distribution for RunB reached $H = 2.3219$ bits, matching the theoretical maximum
$\log_2(5)$ to four decimal places.

Together, these results support the conclusion
that the scheme is a viable source of cooperative, verifiable on-chain randomness
for lottery applications, without any dependency on trusted third parties or
external oracle services.

\section{Security Analysis}
\label{security}

  An attack surface that cannot be fully closed by the deadline block scheme alone is a block producer with sufficient influence over block construction. Such an adversary could selectively include or exclude reveal transactions to steer the XOR aggregate, or in proof-of-stake contexts attempt to manipulate the hash of the reveal deadline block.
  
  Still, one mitigation strategy may be considered: using hash chaining, where the seed incorporates hashes of multiple consecutive blocks rather than a single one. This would raise the cost of a sustained manipulation attack or reduce the time frame to produce a suitable deadline block. Thus, it would make the attack increasingly harder to coordinate and pay for.

\section{Conclusion and Future Work}   
\label{conc}
    
  This work presented a trustless, oracle-free randomness scheme for blockchain-based lotteries, built on a commit-reveal protocol extended with two key contributions: \textit{reveal-as-a-choice} and \textit{block entropy anchoring}. 
  
  Reveal-as-a-choice eliminates the classical denial-of-service vulnerability of commit-reveal schemes. This also removes the need for severe financial penalties or security deposits, lowering the barrier to honest participation without sacrificing the liveness of the application. Furthermore, limiting the number of revealers, possibly rewarded, increases their competition and adds uncertainty to the outcome.
  
  Block entropy anchoring also addresses the last-revealer problem by incorporating the hash of a future block (one that has not yet been mined at the time reveal decisions must be made) as an implicit final contributor to the randomness seed. Because this value is unknown to all participants until after the reveal deadline, no participant can evaluate the two branches of their reveal decision, closing the strategic advantage that the last revealer would otherwise hold.     

  Several directions remain open for the further development of the proposed system.
  The decentralized application (DApp) that serves as the primary interface of the platform is currently in active development.
  The focus is on delivering a complete user experience that combines all contract features and minimizes gas consumption across the contract suite so that the system remains economically viable as network usage grows. 
  
  The current implementation does not support lottery cancelation or ticket refunds, a limitation inherent to the immutable nature of on-chain state. Introducing an \textit{opt-in} cancelation mechanism (for instance, triggered when a minimum participation threshold is not reached before the deadline) would bring the platform closer to parity with traditional lottery systems and reduce the financial risk for participants.
  
\subsubsection{\ackname}
This work is funded by the European Union / Next Generation EU through the project ``Descentralizar Portugal com Blockchain'' (01/C05-i11/2024.PC644918095-00000033) of ``Programa de Recuperação e Resiliência'' (PRR).

%
%
%
\bibliographystyle{splncs04}
\bibliography{bibiography}

@misc{do2024randao,
  author       = {Hai Son Do and Tran, Cong and Dang, Thien and Tran, Quang Huy},
  title        = "{Preventing the Last Revealer Attack on RANDAO with Shamir Secret Sharing}",
  year         = {2024},
  eprint       = {2403.09541},
  archivePrefix = {arXiv},
  primaryClass = {cs.CR},
  url          = {https://arxiv.org/abs/2403.09541},
  note         = {arXiv:2403.09541 [cs.CR]}
}

@inproceedings{algorand,
  author    = {Gilad, Yossi and Hemo, Rotem and Micali, Silvio and Vlachos, Georgios and Zeldovich, Nickolai},
  title     = {{Algorand: Scaling Byzantine Agreements for Cryptocurrencies}},
  booktitle = {26th ACM Symp. on Operating Systems Principles},
  year      = {2017},
  note      = {{DOI}: \href{https://doi.org/10.1145/3132747.3132757}{10.1145/3132747.3132757}}
}

@misc{chainlinkvrf,
  author       = {Breidenbach, Lorenz and Cachin, Christian and Chan, Benedict and Coventry, Alex and Ellis, Steve and Juels, Ari and Koushanfar, Farinaz and Miller, Andrew and Magauran, Brendan and Moroz, Daniel and Nazarov, Sergey and Topliceanu, Alexandru and Tramer, Florian and Zhang, Fan},
  title        = {{Chainlink VRF: On-chain Verifiable Randomness}},
  year         = {2021},
  url          = {https://research.chain.link/whitepaper-v2.pdf},
  howpublished = {Chainlink Technical Whitepaper v2.0}
}

@inproceedings{blum82,
  author    = {Manuel Blum},
  title     = {{Coin Flipping by Telephone: A Protocol for Solving Impossible Problems}},
  booktitle = {24th IEEE Computer Conference (CompCon)},
  year      = {1982},
  note      = {{DOI} \href{https://doi.org/10.1145/1008908.1008911}{10.1145/1008908.1008911}}
}

@techreport{CMU-ISRI-04-126,
    author       = {Latanya Sweeney and Michael I. Shamos},
    title        = "{A Multiparty Computation for Randomly Ordering Players and Making Random Selections}",
    institution  = {Carnegie Mellon University, School of Computer Science, Institute for Software Research International (ISRI)},
    number       = {CMU-ISRI-04-126},
    year         = {2004},
    address      = {Pittsburgh, PA, USA},
    url          = {http://reports-archive.adm.cs.cmu.edu/anon/anon/usr/ftp/isri2004/CMU-ISRI-04-126.pdf}
}

@article{rabin83,
  author    = {Rabin, Michael O.},
  title     = {{Transaction Protection by Beacons}},
  journal   = {Journal of Computer and System Sciences},
  year      = {1983},
  volume    = {27},
  number    = {2},
  note      = {{DOI}: \href{https://doi.org/10.1016/0022-0000(83)90042-9}{10.1016/0022-0000(83)90042-9}}
}

@inproceedings{rabin99,
  author    = {Micali, Silvio and Rabin, Michael and Vadhan, Salil},
  title     = {{Verifiable Random Functions}},
  booktitle = {40th IEEE Annual Symp. on Foundations of Computer Science},
  year      = {1999},
  address   = {New York, NY, USA},
  note      = {{DOI}: \href{https://doi.org/10.1109/SFFCS.1999.814584}{10.1109/SFFCS.1999.814584}}
}

@article{ayton04,
  author  = {Ayton, Peter and Fischer, Ilan},
  title   = {{The hot hand fallacy and the gambler's fallacy: Two faces of subjective randomness?}},
  journal = {Memory \& Cognition},
  year    = {2004},
  volume  = {32},
  number  = {8},
  note    = {{DOI}: \href{https://doi.org/10.3758/BF03206327}{10.3758/BF03206327}}
}

@article{gilovich85,
  author       = {Gilovich, Thomas and Vallone, Robert and Tversky, Amos},
  title        = {{The hot hand in basketball: On the misperception of random sequences}},
  journal      = {Cognitive Psychology},
  volume       = {17},
  number       = {3},
  year         = {1985},
  note         = {{DOI}: \href{https://doi.org/10.1016/0010-0285(85)90010-6}{10.1016/0010-0285(85)90010-6}}
}

@inproceedings{bonneau15,
  author       = {Bonneau, Joseph and Miller, Andrew and Clark, Jeremy and Narayanan, Arvind and Kroll, Joshua A. and Felten, Edward W.},
  title        = {{SoK: Research Perspectives and Challenges for Bitcoin and Cryptocurrencies}},
  booktitle    = {IEEE Symp. on Security and Privacy},
  year         = {2015},
  note         = {{DOI}: \href{https://doi.org/10.1109/SP.2015.14}{10.1109/SP.2015.14}}
}

@book{david62,
  author    = {David, Florence Nightingale},
  title     = {{Games, Gods and Gambling: A History of Probability and Statistical Ideas}},
  publisher = {Dover Publications},
  year      = {1962},
  note      = {{ISBN-13} 978-0486400235}
}

@article{kahneman72,
  author  = {Kahneman, Daniel and Tversky, Amos},
  title   = {{Subjective probability: A judgment of representativeness}},
  journal = {Cognitive Psychology},
  year    = {1972},
  volume  = {3},
  number  = {3}
}

@article{croson05,
  author  = {Croson, Rachel and Sundali, James},
  title   = {{The Gambler's Fallacy and the Hot Hand: Empirical Data from Casinos}},
  journal = {Journal of Risk and Uncertainty},
  year    = {2005},
  volume  = {30},
  number  = {3},
  note    = {{DOI}: \href{https://doi.org/10.1007/s11166-005-1153-2}{10.1007/s11166-005-1153-2}}
}

@book{mlodinow08,
  author    = {Mlodinow, Leonard},
  title     = {{The Drunkard's Walk: How Randomness Rules Our Lives}},
  publisher = {Pantheon Books},
  year      = {2008},
  note      = {{ISBN} 978-0-375-4204-5}
}

@misc{rfc:9381,
  author="S. Goldberg and L. Reyzin and D. Papadopoulos and J. Včelák",
  title={{Verifiable Random Functions (VRFs)}},
  howpublished="RFC 9381 (Informational)",
  series="Internet Request for Comments",
  year=2023,
  month=aug,
  note="{DOI: \href{https://doi.org/10.17487/RFC9381}{10.17487/RFC9381}}"
}

@article{Matsumoto98,
  title = {{Mersenne Twister: A 623-Dimensionally Equidistributed Uniform Pseudo-Random Number Generator}},
  author = {Makoto Matsumoto and Takuji Nishimura},
  journal = "ACM Trans. on Modeling and Computer Simulation",
  volume = "8",
  number = "1",
  month = jan,
  year = {1998},
  pages = {3–30},
  note = "{DOI: \href{https://doi.org/10.1145/272991.272995}{10.1145/272991.272995}}"
}

@article{brassard1988,
  author    = {Brassard, Gilles and Chaum, David and Cr{\'{e}}peau, Claude},
  title     = {Minimum Disclosure Proofs of Knowledge},
  journal   = {Journal of Computer and System Sciences},
  volume    = {37},
  number    = {2},
  pages     = {156--189},
  year      = {1988},
  doi       = {10.1016/0022-0000(88)90005-0}
}

@inproceedings{boneh2018vdf,
  author    = {Boneh, Dan and Bonneau, Joseph and B{\"{u}}nz, Benedikt and Fisch, Ben},
  title     = {Verifiable Delay Functions},
  booktitle = {Advances in Cryptology -- {CRYPTO} 2018},
  series    = {Lecture Notes in Computer Science},
  volume    = {10991},
  pages     = {757--788},
  publisher = {Springer},
  year      = {2018},
  doi       = {10.1007/978-3-319-96884-1\_25},
  note      = {ePrint: \url{https://eprint.iacr.org/2018/601}}
}

@inproceedings{yakira2020,
  author    = {Yakira, David and Grayevsky, Ido and Asayag, Avi and Keidar, Idit},
  title     = {Economically Viable Randomness},
  booktitle = {Proceedings of the 2nd {ACM} Conference on Advances in
               Financial Technologies ({AFT} 2020)},
  pages     = {133--145},
  year      = {2020},
  publisher = {ACM},
  doi       = {10.1145/3419614.3423249},
  note      = {arXiv: \url{https://arxiv.org/abs/2007.03531}}
}

@misc{lee2025,
  author       = {Lee, Suhyeon and Gee, Euisin},
  title        = {Commit-Reveal\textsuperscript{2}: Securing Randomness Beacons with Randomized Reveal Order in Smart Contracts},
  year         = {2025},
  eprint       = {2504.03936},
  archivePrefix= {arXiv},
  primaryClass = {cs.CR},
  url          = {https://arxiv.org/abs/2504.03936}
}

@article{hassan2023,
  author    = {Hassan, Abid and others},
  title     = {From Trust to Truth: Advancements in Mitigating
               the Blockchain Oracle Problem},
  journal   = {Journal of Network and Computer Applications},
  volume    = {217},
  pages     = {103672},
  year      = {2023},
  publisher = {Elsevier},
  doi       = {10.1016/j.jnca.2023.103672}
}

@article{chompurmpakdee2023,
  author    = {Chompurmpakdee, Napat and others},
  title     = {{NativeVRF}: A Simplified Decentralized Random Number
               Generator on {EVM} Blockchains},
  journal   = {Systems},
  volume    = {11},
  number    = {7},
  pages     = {326},
  year      = {2023},
  publisher = {MDPI},
  doi       = {10.3390/systems11070326}
}
\end{document}